\documentclass[useAMS,usenatbib,referee]{mn2e}

\usepackage{amsmath}
\usepackage{amssymb}
\usepackage{graphicx}
\usepackage{dcolumn}
\usepackage{bm}
\usepackage{ulem}

\title[Anisotropic crystal structure of magnetized neutron star crust]
{Anisotropic crystal structure of magnetized neutron star crust}
\author[D.A. Baiko and A.A. Kozhberov]
{D.A. Baiko\thanks{E-mail:baiko@astro.ioffe.ru} and A.A. Kozhberov \\
Ioffe Institute,
Politekhnicheskaya 26, 194021 Saint Petersburg, Russian Federation}

\begin{document}

\date{Accepted; Received ; in original form}

\pagerange{\pageref{firstpage}--\pageref{lastpage}} \pubyear{2013}

\maketitle

\begin{abstract}
%
Although crystallized neutron star crust is responsible 
for many fascinating observational phenomena, its actual microscopic 
structure in tremendous gravitational and magnetic fields is not 
understood. Here we show that in a non-uniform magnetic field, 
three-dimensional ionic Coulomb crystals comprising the crust may  
stretch or shrink while their electrostatic pressure becomes 
anisotropic. The pressure depends non-linearly on the magnitude of the 
stretch, so that a continuous magnetic field evolution may result in 
an abrupt crystal elongation or contraction. This may provide 
a trigger for magnetar activity. A phonon mode instability is revealed, 
which sets the limits of magnetic field variation beyond which the 
crystal is destroyed. These limits sometimes correspond to surprisingly 
large deformations. It is not known what happens to crust matter 
subject to a pressure anisotropy exceeding these limits. We hypothesize 
that the ion system then possesses a long-range order only in one or 
two dimensions, that is becomes a liquid crystal.
%
\end{abstract}


\section{Introduction}
Neutron stars are magnificent astrophysical objects for a wide range
of stunning observational phenomena associated with them 
as well as for extraordinary challenges and opportunities they present 
to a theorist trying to explain their behavior \citep[e.g.,][]{K10}. 
Among the most striking manifestations of neutron stars, one 
can mention pulsars, amazing machines converting the star rotation 
energy into multi-wavelength highly periodic radiation and accelerated 
particles; millisecond pulsars, the most accurate clocks in the 
Universe; soft-gamma repeaters, famous for their restless bursting 
activity and mind-boggling giant flares. 

From a theorist's point of view, the most exciting is probably the core 
of a neutron star, which is compressed to super-nuclear densities by 
immense gravity. Essentially, this is a unique laboratory of strong 
interaction physics \citep[e.g.,][]{HPY07}. The core is hidden from us 
by an $\sim 1$ km thick crust which may be threaded by magnetic fields 
of up to at least $\sim 10^{15}$ G. This underscores the need for 
a microscopic model of magnetized neutron star crust to enable 
extraction of information coming from the core. Of great theoretical 
interest is also the origin of such colossal fields, whereas their 
interaction and joint evolution with the crust are believed to 
be chiefly responsible for the observed phenomena.  

The density $\rho$ of the crust matter spans many orders 
of magnitude. One usually distinguishes the outer crust at 
$\rho < 4.3 \times 10^{11}$ g cm$^{-3}$ and the inner crust at densities 
above that and up to $\sim 1.5 \times 10^{14}$ g cm$^{-3}$ where the 
core starts. The bulk of the outer crust consists of fully ionized 
atomic nuclei of various sorts and strongly degenerate nearly 
incompressible electron gas. In the inner crust in addition to ions and 
electrons there are degenerate neutrons dripped from the nuclei. In view 
of a rapid gravitational separation, it is typically assumed that at any 
given density in the crust there is only one ion species. 
If any possible but not firmly established effects associated 
with dripped neutrons are neglected, such 
electron-ion system is governed by pure Coulomb forces and is known to 
crystallize into a Coulomb crystal at a sufficiently low 
temperature \citep{NNN87}. This occurs at very early stages of the 
neutron star life across the entire crust with the exception of the 
outermost low-density layers. 

It is well-known that the body-centered cubic (bcc) lattice has 
the lowest electrostatic energy among all one-component Coulomb 
crystals with uniform electron background, however, the energy 
difference between several stable structures is tiny. Previous 
studies of Coulomb solids have thus focused on a handful of lattices 
characterized by near minimum energies and with a possible presence of 
a uniform magnetic field \citep[e.g.,][]{CK55,C61,NF82,NF83,
BSO92,BPY01,KB15,CF16}. Formation of the 
bcc crystal in neutron star envelopes has been confirmed in recent 
molecular dynamics simulations \citep[][]{EYC16}.
Linear elasticity theory based on the bcc lattice has been used to 
study neutron star crust oscillations \citep[e.g.,][]{P05,S16}. 
Molecular dynamics simulations of bcc crystal breaking have been 
performed \citep[e.g.,][]{HK09}. However, as we demonstrate below, 
these approaches miss a significant part of the story.

\section{Anisotropic crystal pressure}
\label{pressure}
Consider a bcc Coulomb crystal of ions with rigid and uniform electron 
background in the outer neutron star crust. Suppose the crystal is 
stretched or shrunken in some direction by a stretch factor $\xi$, 
which means that projections of all lattice vectors on this direction 
are multiplied by $\xi$ while the overall scale factor is adjusted to 
maintain constant ion number density. The system then remains a perfect 
crystal. If one neglects the ion motion about the lattice nodes and the 
electron polarisation which are typically valid zero-order assumptions, 
the system energy $U(\xi)$ is the sum of the energy of strongly 
degenerate electron gas and the electrostatic (Madelung) energy of ions. 

Let us denote lattice vectors of the stretched crystal as $\bm{R}$
and apply a uniform infinitesimal deformation 
$R_\alpha \to R_\alpha + u_{\alpha \beta} R_\beta$. Then, to first 
order in $u_{\alpha \beta}$, the energy of the deformed crystal is 
%
\begin{equation}
      U(\xi,u_{\alpha \beta}) = U(\xi) + V \sigma_{\alpha \beta}(\xi) 
      u_{\alpha \beta}~,
\end{equation}
%
where $V$ is the volume and $\sigma_{\alpha \beta}(\xi)$ is the stress 
tensor of the crystal stretched by $\xi$. The stress tensor consists
of two terms
%
\begin{equation}
    \sigma_{\alpha \beta} = - P^{\rm e} \delta_{\alpha \beta}
    + S^{\rm st}_{\alpha \beta}~,     
\end{equation}
%
where $P^{\rm e}$ is the degenerate electrons' pressure 
\citep[e.g.,][]{LL80,PY12} and $S^{\rm st}_{\alpha \beta}$
is the electrostatic (ion) contribution. For bcc lattice the latter 
contribution is isotropic 
$S^{\rm st}_{\alpha \beta}(\xi=1) = - 
P^{\rm st}_b \delta_{\alpha \beta}$. The electrostatic pressure
of the bcc lattice \citep{F36} is  
$P^{\rm st}_b = - \zeta n Z^2 e^2 /(3 a) < 0$, where 
$Z|e|$ is the ion charge, $n$ is the ion number density, 
$\zeta \approx 0.895929255682$ is 
the Madelung constant of the bcc lattice, and
$a=(4\pi n/3)^{-1/3}$ is the ion sphere radius.

\begin{figure}                                           
\begin{center}                                              
\leavevmode                                                 
\includegraphics[height=73mm,bb=70 533 541 744,clip]{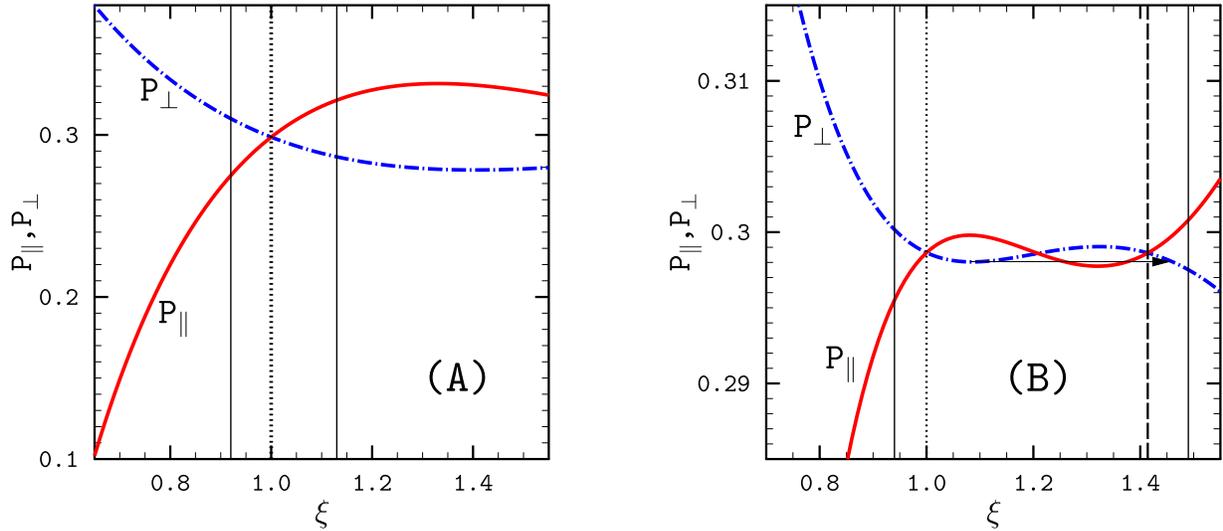} 
\end{center}                                                
\vspace{-0.4cm} 
\caption[]{Components of the ion pressure tensor in 
units of $-nZ^2e^2/a$ versus the stretch 
factor $\xi$. The data are plotted for the bcc lattice stretched along 
the direction towards the nearest neighbour or the cube diagonal (A) 
and along the direction towards the second nearest neighbour or the cube 
edge (B). Solid (red) line is the pressure along the stretch direction 
and dot-dashed (blue) line is the pressure in the perpendicular 
direction. $\xi=1$ (dots) and $\xi=\sqrt{2}$ (dashes) indicate the bcc 
and fcc lattices, respectively. Thin solid vertical lines mark 
the critical stretch factors at which the phonon mode instability first 
occurs.}                                             
\label{press}
\end{figure}

For $\xi \ne 1$, one can calculate the ion contribution to the stress 
tensor exactly using a practical formula \citep[][]{F36,B11}. 
The pressure is no longer isotropic.
In Fig.\ \ref{press}, panel (A), we stretch the crystal along the 
direction towards the nearest neighbour i.e. along the {\it diagonal} 
of the main lattice cube. In panel (B) the stretch is along
the direction towards the second nearest neighbour i.e. along the 
{\it edge} of the main lattice cube. In both cases the 
pressure is isotropic in the plane perpendicular to the stretch 
direction. We denote it $P_\perp$ and show by dot-dashed (blue) lines 
in Fig.\ \ref{press}. The pressure along the stretch direction, 
$P_\parallel$, is shown by solid (red) curves and is clearly not the 
same as $P_\perp$. The dependence of pressure components on $\xi$ is 
seen to be non-linear, especially in panel (B), which limits the 
accuracy of a standard elastic expansion starting at $\xi=1$.        
  
%

The stretch in panel (B) at $\xi = \sqrt{2}$ is known 
\citep[e.g.,][]{NF82} to produce a face-centered cubic (fcc) 
lattice, which is isotropic. The stretch in panel (A) also produces 
isotropic structures: a simple cubic lattice at $\xi=2$ and the fcc 
lattice at $\xi=4$. 

It is worth noting that the two stretch directions considered coincide 
with the highest symmetry axes of the bcc crystal. 
Stretching the crystal along the direction 
towards the third nearest neighbour i.e.\ along the {\it diagonal
of the face} of the main lattice cube yields the pressure 
tensor with three different eigenvalues. Even though further work is 
desirable to render a more complete understanding of Coulomb crystals' 
behavior under stretches, cases (A) and (B) will be sufficient for
our discussion. Also, other types of deformations (e.g., shear) can be 
studied in a similar fashion.

\section{Equilibrium in non-uniform magnetic field}
\label{hydro}
External forces of various nature can cause crystal stretching just 
discussed. One possibility, which seems to be especially relevant for
neutron star crusts, is the action of a non-uniform magnetic field. 
For illustration of expected effects, let us construct a 
microscopic model of local hydrostatic equilibrium of crystallized 
neutron star crust matter in a stationary magnetic field 
$\bm{B}(\bm{r})$. The momentum conservation equation 
\citep[e.g.,][]{Cetal86} can be written as      
%
\begin{equation}
   \frac{\partial}{\partial r_\beta} (\sigma_{\alpha \beta} + 
   \tau_{\alpha \beta}) - \rho \frac{\partial \psi}{\partial r_\alpha} 
   = 0~,
\label{hydroeq}
\end{equation}
%
where $\tau_{\alpha \beta} = 
(2 B_\alpha B_\beta - B^2 \delta_{\alpha \beta})/(8\pi)$,
$- \nabla \psi = \bm{g}$ is the gravitational acceleration, and $\rho$ 
is the mass density.

Of primary interest for us are magnetic field configurations which 
cannot be in hydrostatic equilibrium with matter characterized by 
isotropic pressure, i.e.\ which cannot be stationary in a liquid star. 
Such configurations may be produced upon freezing 
of a non-stationary liquid system or be a result of magnetic field 
evolution already in the solid phase. 

Consider the simplest geometry of magnetic field and gravity: 
$\bm{B} = B(x) \bm{e}_z$ and $\bm{g} = - g \bm{e}_z$ ($g$ is constant), 
for which there is no hydrostatic equilibrium in a liquid. To prove 
the latter statement we observe that in this case 
$-\tau_{xx} = -\tau_{yy} = \tau_{zz} = B^2/(8\pi)$, and
if the total pressure $P$ is strictly isotropic, 
$\sigma_{xx} = \sigma_{yy} = \sigma_{zz} = - P(\rho)$. 
Then Eq.\ (\ref{hydroeq}) reduces to    
%
\begin{eqnarray}
     \frac{\partial P}{\partial z} &=& - g \rho~,
\label{isoz}\\
     \frac{\partial P}{\partial x} &=& 
   - \frac{1}{8 \pi} \, \frac{\partial B^2}{\partial x}~.
\label{isox}
\end{eqnarray}
%
Equation (\ref{isox}) implies that $P(\rho) = f(z) - B^2(x)/(8 \pi)$, 
where $f(z)$ is some function, and thus $\rho$ depends on both,
$x$ and $z$. However, Eq.\ (\ref{isoz}) can be written as 
$\rho = - f'(z)/g$ which is independent of $x$. The contradiction
means that the stress tensor of matter must be anisotropic and the 
easiest way to realize this is to impose a crystal stretch.   

The stretch direction and magnitude depend on specific boundary 
conditions. Here we assume, somewhat arbitrarily, that the stretch 
is along the $z$-axis and $\xi(z,x=0)=1$. 
Then $\sigma_{zz} = - P_\parallel - P^{\rm e}$ and
$\sigma_{xx} = - P_\perp - P^{\rm e}$, where
$P_{\parallel,\perp}$ are determined by both mass density and the 
stretch factor, while $P^{\rm e}$ is a function of $\rho$ only.  
Consequently, Eq.\ (\ref{hydroeq}) can be rewritten as
%
\begin{eqnarray}
     \frac{\partial}{\partial z} (P^{\rm e} + P_\parallel) &=& 
     - g \rho~,
\label{hydroz} \\
     \frac{\partial}{\partial x} (P^{\rm e} + P_\perp) &=& 
   - \frac{1}{8 \pi} \, \frac{\partial B^2}{\partial x}~.
\label{hydrox}
\end{eqnarray}
%

Let us introduce functions $\Delta_{\parallel,\perp}(\xi)$ as 
$P_{\parallel,\perp} = [1+\Delta_{\parallel,\perp}(\xi)] 
P^{\rm st}_b(\rho)$ so that $\Delta_{\parallel,\perp}(1)=0$
(cf.\ Fig.\ \ref{press}). 
At $x=0$, Eq.\ (\ref{hydroz}) reduces to the equation of hydrostatic 
equilibrium in the absence of the magnetic field:
%
\begin{equation}
    \frac{\partial [P^{\rm e}(z,0)+P^{\rm st}_b(z,0)]}{\partial z}  
    = - g \rho(z,0)~.
\end{equation}
%
Integrating Eq.\ (\ref{hydrox}) we obtain
%
\begin{equation}
  \delta P_\rho(z,x) + P^{\rm st}_b(z,x) \Delta_{\perp}(\xi(z,x))  
   = \frac{B^2(0) - B^2(x)}{8\pi}~,
\label{compens}
\end{equation}
%
where
%
\begin{equation}
 \delta P_\rho (z,x) = P^{\rm e}(z,x) + P^{\rm st}_b(z,x) 
                - P^{\rm e}(z,0) - P^{\rm st}_b(z,0)
\end{equation}
%
is the deviation of the isotropic contribution to the pressure from
its field-free value at given $z$. The two terms on the left-hand 
side of Eq.\ (\ref{compens}) thus express the fact that the magnetic 
pressure is compensated by a combined action of a mass density change
and a crystal stretch.

To take this consideration a step further, we assume that 
$\Delta_{\parallel}(\xi) = - C \Delta_{\perp}(\xi)$, where $C \sim 1$ 
is a positive constant. In particular, $\Delta_{\parallel}$ and 
$\Delta_{\perp}$ are proportional to each other for a linear dependence 
of $P_{\parallel,\perp}-P^{\rm st}_b$ on $\xi-1$ 
(cf.\ Fig.\ \ref{press}). Then we can differentiate Eq.\ (\ref{compens}) 
with respect to $z$ and use the derivative of the second term on the 
left-hand side to express the derivative of $P_\parallel$ in 
Eq.\ (\ref{hydroz}). This results in
\begin{equation}
    \frac{\partial \delta P_\rho}{\partial z} = 
    - \frac{g}{1+C} \delta \rho~,
\label{zfin}
\end{equation}
where $\delta \rho(z,x) = \rho(z,x)-\rho(z,0)$. Suppose, for certainty,
that $B^2(x)<B^2(0)$. Then $\delta \rho, \delta P_\rho \ge 0$ and since
$P^{\rm st}_b<0$, $\Delta_{\perp} \le 0$, i.e. $\xi \ge 1$ 
(cf.\ Fig.\ \ref{press}). According to Eqs.\ (\ref{zfin}) and 
(\ref{compens}), $\delta P_\rho$ decreases with increase of $z$, 
whereas $\xi$ increases, so that the relative importance of the first 
term on the left-hand side of Eq.\ (\ref{compens}) diminishes
with $z$. The actual value of $\delta P_\rho$  
depends on the assumed boundary condition for $\delta \rho$ at a 
fixed $z$. 

If $\delta \rho = 0$ at some $z_0$ it will remain zero 
at $z>z_0$ and so will $\delta P_\rho$. In this situation, $\rho$, 
$P^{\rm e}$, and $P^{\rm st}_b$ are independent of $x$ and the magnetic 
pressure is balanced solely by the crystal stretch: 
%
\begin{equation}
    \Delta_\perp(\xi(z,x)) = 
     \frac{B^2(0) - B^2(x)}{8\pi P^{\rm st}_b(z)}~.   
\label{anizo}
\end{equation}
%
The stretch factor $\xi(z,x)$ can be found by applying a well-defined
function $\Delta_\perp^{-1}$ to this formula. 
Equation (\ref{anizo}) represents the maximum effect of the magnetic 
field on the crystal shape and will be adopted for the sake of the 
qualitative discussion in the following Section.

\section{Limits of crystal anisotropy}
Given a Coulomb crystal with an arbitrary lattice,
one can calculate its dynamic matrix using the standard 
formulas \citep{CK55,B02} and analyse the crystal phonon modes. A bcc 
Coulomb crystal ($\xi=1$) has three phonon modes at each wavevector 
$\bm{k}$ in the first Brillouin zone, two of which are transverse 
acoustic, while the third one is longitudinal optic. 

If $\xi \ne 1$ but is close to 1, the same picture persists. However,
we have found that as the stretch or contraction increased, 
one of the acoustic modes developed an instability 
manifested by the appearance of imaginary frequencies at certain 
isolated directions of $\bm{k}$ in the vicinity 
of $k=0$ (i.e., at large wavelengths). In general,
such an instability signifies a crystal destruction. The critical $\xi$ 
(denoted $\xi_{\rm crit}$) are marked by
thin solid vertical lines in Fig.\ \ref{press}. 

In Fig.\ \ref{press}(A) we observe a weakly non-linear dependence of 
the pressure tensor components on $\xi$ and the instability occurs at 
$\xi_{\rm crit} \approx 0.92$ or 1.13 when the anisotropy reaches 
$\sim 10\%$. Equation (\ref{anizo}) applied at $\xi_{\rm crit}$ 
enables one to estimate magnetic field variation $(\delta B)_{\rm crit}$ 
which can be supported by the crystal as
%
\begin{equation}
    B (\delta B)_{\rm crit} \approx 2.2 \times 10^{21} \, 
    \Delta_{0.1} \, \rho^{4/3}_6 Z_{26}^2 A_{56}^{-4/3} \,\, 
    [{\rm G}^2]~,
\label{BdB} 
\end{equation}
%
where we have used the explicit expression for $P^{\rm st}_b$, 
$\Delta_{0.1} \equiv \Delta_\perp(\xi_{\rm crit})/0.1$, $\rho_6$ 
is the mass density in units of $10^6$ g cm$^{-3}$, 
$Z_{26} \equiv Z/26$, $A_{56} \equiv A/56$, and $A$ is the ion mass 
number. Hence, for iron at $10^6$ g cm$^{-3}$ and a typical 
pulsar field of $10^{12}$ G, 
$(\delta B)_{\rm crit} / B \sim 2 \times 10^{-3}$. 
For a typical magnetar field of $4 \times 10^{14}$ G, 
$(\delta B)_{\rm crit} / B \sim 10^{-8}$. Assuming a specific magnetic 
field geometry, one can deduce the maximum size of a perfect crystal, 
which may be present in such matter, as the distance over which 
magnetic field varies by $(\delta B)_{\rm crit}$. For 
instance, in the geometry where $\delta B \sim B$ over the star's 
radius $\sim 10^6$ cm, there may be $\sim 20$ m crystals in pulsars 
but only $\sim 0.1$ mm crystals in magnetars at $\rho=10^6$ g cm$^{-3}$.
If there is a ``spot'' with $\delta B \sim B$ over $10^4$ cm, then 
the crystals available will be 100 times smaller.    
 
At higher densities a perfect crystal can withstand much stronger 
field variations. If the dripped neutrons in the inner crust 
simply add another isotropic term to the total stress, setting 
$\rho=10^{14}$ g cm$^{-3}$, $Z=40$, and $A=1000$, we obtain 
$B (\delta B)_{\rm crit} \approx 5.3 \times 10^{30}$ G$^2$. It follows 
that $(\delta B)_{\rm crit} \sim B$ can be accommodated by the perfect 
crust at $B \lesssim 2 \times 10^{15}$ G. Incidentally, this is the same 
number as the maximum magnetic field presently reported for magnetars 
\citep{OK14} and it is much smaller than the field required for 
complete inner crust magnetization \citep{BY13}. 

A surprising result is illustrated in Fig.\ \ref{press}(B). In this 
case the instability occurs very quickly upon a crystal contraction
($\xi_{\rm crit} \approx 0.94$), 
but if the crystal is elongated, the pressure stays more or less 
isotropic (anisotropy does not exceed $2\%$) and the structure remains 
stable up to $\xi_{\rm crit} \approx 1.5$. The range of stable 
structures extends all the way to the fcc lattice and beyond. This is 
in contrast to Fig. \ref{press}(A), where the crystal would be 
destroyed before reaching fcc at $\xi=4$.

From this picture it appears that the instability is not caused by 
the stretch itself but by the pressure anisotropy associated with it.
The value of the available strain is remarkable being several times 
greater than the typical value of $0.1$, obtained by molecular dynamics 
simulations of breaking strain in a shearing crystal 
\citep{HK09,CH10,CH12,JO13}, even though the respective pressure 
anisotropy is very weak. It would require 
$B (\delta B)_{\rm crit}$ approximately 10 times smaller than the 
estimate (\ref{BdB}) in view of different $y$-axis scales in 
Figs. \ref{press}(A) and (B). 
Comparing critical stresses and strains in Figs.\ \ref{press}(A)
and (B) one notices their pronounced dependence on the stretch 
direction. This calls into doubt the applicability of 
phenomenological von Mises and Tresca criteria for crystal breaking
often used in neutron star crust contexts \citep[e.g.,][]{UCB00}.    

Suppose that the crystal is oriented in such a way that a magnetic 
field variation results in an increase of $\xi$ in the direction
analysed in panel (B). This can continue till the minimum of 
$-P_\perp$ is reached but any further variation of $B$ in the same 
direction (e.g., as a result of field drift or decay) would require 
a jump of the stretch factor shown schematically in panel (B) by an 
arrow. Such a strong jump from $\xi \approx 1.08$ to $\approx 1.46$ 
coupled with a structural transition from a stretched bcc to 
a stretched fcc lattice might provoke a 
dramatic crust rearrangement, which could affect the stability of 
neighbouring crust regions and serve as a trigger mechanism for 
magnetar activity. 

\begin{figure}                                           
\begin{center}                                              
\leavevmode                                                 
\includegraphics[height=73mm,bb=72 532 540 740,clip]{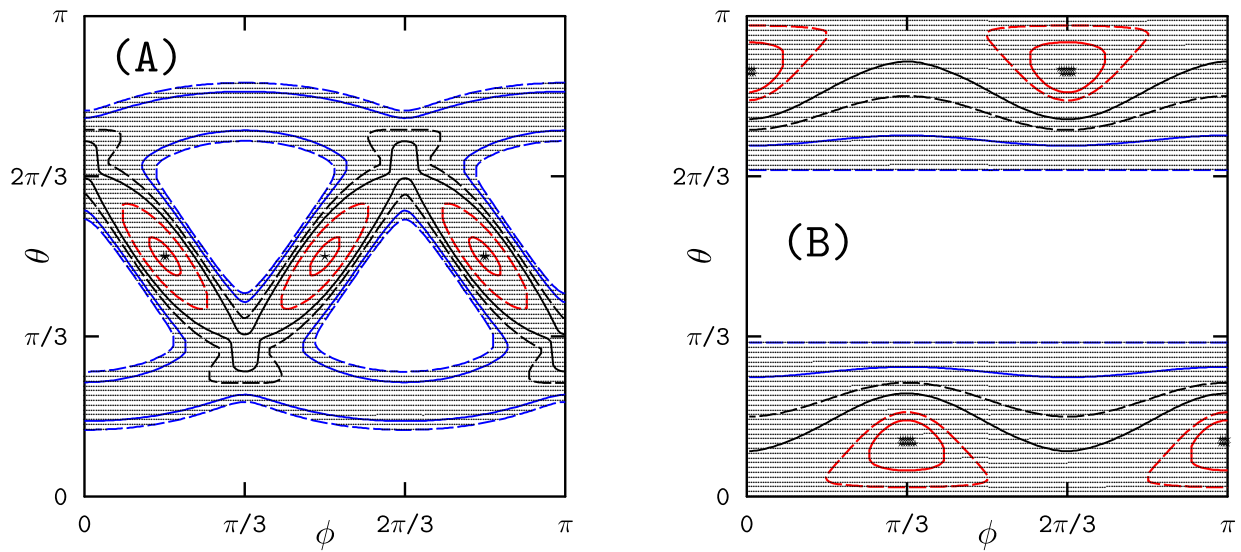} 
\end{center}                                                
\vspace{-0.4cm} \caption[]{Ranges of propagation angles of unstable 
phonon modes. $\theta$ and $\phi$ are spherical angles in a 
reference frame whose $z$-axis is directed along the stretch. Stars 
show the very onset of the instability at $\xi=1.14$ (A) or 
$\xi=0.918$ (B). Sequentially expanding contours encircle unstable 
regions at $\xi=1.15, 1.2, 1.25, 1.3, 1.4, 1.5$ (A) or 
$\xi=0.91, 0.9, 0.85, 0.8, 0.7, 0.5$ (B). The regions at $\xi=1.5$ (A) 
and $\xi=0.5$ (B) are highlighted by shading.}                                             
\label{zones}
\end{figure}
%

A natural question arises: what happens with the crust if the
pressure anisotropy exceeds the values corresponding to
$\xi_{\rm crit}$? It is not excluded that another three-dimensional 
crystal structure is stronger than the bcc lattice but a really 
substantial improvement seems unlikely. In principle, at low 
temperatures and strongly anisotropic pressures, the Coulomb system 
can be in a new, unknown state. To investigate this possibility we 
have performed a numerical experiment. If we artificially take $\xi$ 
in the crystal dynamic matrix formula beyond its critical value, the 
phonon instability will spread over finite ranges of propagation angles 
and to shorter wavelengths. This is illustrated in Fig.\ \ref{zones} 
where panels (A) and (B) correspond to super-critical crystal 
elongation and contraction, respectively, in the direction of the 
nearest neighbour [as in Fig.\ \ref{press}(A)]. The propagation angles 
are defined in a spherical reference frame with the $z$-axis in the 
direction of the stretch. Stars represent the unstable directions at 
$\xi=1.14$ (A) or $\xi=0.918$ (B), which is very close to the 
instability onset [cf.\ Fig.\ \ref{press}(A)]. 
It turns out, that in panel (A) these are directions towards
the third nearest neighbours in the plane perpendicular to the stretch,
but in panel (B) they do not coincide with any obvious axes of
symmetry.
Sequentially expanding contours encircle unstable regions at 
$\xi=1.15, 1.2, 1.25, 1.3, 1.4, 1.5$ in panel (A) 
and at $\xi=0.91, 0.9, 0.85, 0.8, 0.7, 0.5$ in panel (B). Shading 
emphasizes maximum unstable configurations attained in 
Fig.\ \ref{zones}. 

The white regions correspond to propagation angles at which all three
phonon modes remain stable all the way down to $k=0$. In general, these
domains shrink with increase of $|\xi - \xi_{\rm crit}|$. At first, 
they can be described as bundles of planes in $\bm{k}$-space containing 
the origin as in panel (A) for $\xi \lesssim 1.25$ and in panel (B) 
for all $\xi$. At higher stretches they may reduce to bundles
of lines through the origin as in panel (A).

The unstable branches of the phonon spectrum look similar
to collective modes of a Coulomb liquid, which disappear  
at finite $k$ and do not reach $k=0$ \citep[e.g.,][]{SZRT97}. 
This suggests a loss of the long-range order in the respective 
directions in the crystal. The persistence of the long-wavelength 
transverse modes in other directions even for relatively large 
deformations implies, by extension, that the long-range order in one or 
two dimensions may be preserved. Together these observations can be 
taken as an indication that at strongly anisotropic pressure the 
three-dimensional Coulomb system behaves as a liquid crystal.

\section{Discussion}
In neutron star crust, a bcc Coulomb solid will stretch 
or contract in response to a wide class of non-uniform magnetic fields. 
The ensuing crystal will have anisotropic electrostatic pressure, 
which may depend on the stretch factor in a very non-linear fashion. 
In some geometries, this may result in a strongly discontinuous response 
of the crystal structure to a continuous variation of the applied 
(magnetic) stress and serve as a trigger mechanism for magnetar 
activity. The three-dimensional Coulomb crystal develops a phonon mode 
instability (and fails) if the degree of the pressure anisotropy reaches 
$\lesssim 10 \%$ which may correspond to a stretch factor as high as 
$\sim 150 \%$. A respective estimate of the maximum magnetic field 
sustainable by a perfectly crystallized crust is consistent with the 
maximum soft gamma repeater field currently inferred from observations. 
It is unknown what state the crust is in at higher pressure 
anisotropies. One possibility alluded to here is that it resembles a 
liquid crystal having a long-range order in one or two dimensions only. 
Molecular dynamics simulations of Coulomb systems under anisotropic 
pressure may prove instrumental in discovering these structures.  

Anisotropic Coulomb crystals and liquid Coulomb crystals (if exist) 
will have thermodynamic, kinetic, and elastic properties very different  
from those of the standard bcc. Such diverse microscopic quantities as 
neutrino emissivities, diffusion rates, plasma screening factors for 
nuclear reactions will be seriously affected. The strongest effect on 
the neutron star structure is expected near the star's surface, while 
global characteristics, e.g., the mass--radius relation, most likely 
will not change due to the relative smallness of the electrostatic 
pressure.

\section*{Acknowledgments}
We are grateful to A.I.\ Chugunov and D.G.\ Yakovlev for discussions.
A.A.K.\ thanks Leading Science School 9297.2016.2 for support.


\begin{thebibliography}{}

\bibitem[\protect\citeauthoryear{Baiko et al.}
      {Baiko, Potekhin \& Yakovlev}{2001}]{BPY01}
      Baiko D.A.,  Potekhin A.Y., Yakovlev D.G., 2001,   
      Phys.\ Rev.\ E, 64, 057402

\bibitem[\protect\citeauthoryear{Baiko}{2002}]{B02}
      Baiko D.A., 2002, Phys.\ Rev.\ E, 66, 056405, Eq.\ (19)

\bibitem[\protect\citeauthoryear{Baiko}{2011}]{B11}
      Baiko D.A., 2011, MNRAS, 416, 22, Eqs.\ (B1) and (B2) 
       
\bibitem[\protect\citeauthoryear{Baiko \& Yakovlev}{2013}]{BY13}
      Baiko D.A., Yakovlev D.G., 2013, MNRAS, 433, 2018.

\bibitem[\protect\citeauthoryear{Baldereschi et al.}
      {Baldereschi, Senatore \& Oriani}{1992}]{BSO92}
	    Baldereschi A., Senatore G., Oriani I., 1992,
	    Solid State Comm., 81, 21

\bibitem[\protect\citeauthoryear{Carr}{1961}]{C61}
      Carr W.J.Jr., 1961, Phys.\ Rev., 122, 1437

\bibitem[\protect\citeauthoryear{Carroll et al.}{1986}]{Cetal86}
      Carroll B.W., Zweibel E.G., Hansen C.J., McDermott P.N.,
      Savedoff M.P., Thomas J.H., van Horn H.M., 1986, ApJ, 
      305, 767, Eq.\ (8)

\bibitem[\protect\citeauthoryear{Chugunov \& Horowitz}{2010}]{CH10}
      Chugunov A.I., Horowitz C.J., 2010, MNRAS, 407, L54 

\bibitem[\protect\citeauthoryear{Chugunov \& Horowitz}{2012}]{CH12}
      Chugunov A.I., Horowitz C.J., 2012, 
      Contrib.\ Plasm.\ Phys., 52, 122 

\bibitem[\protect\citeauthoryear{Cohen \& Keffer}{1955}]{CK55}
      Cohen M.H., Keffer F., 1955, Phys.\ Rev., 99, 1128

\bibitem[\protect\citeauthoryear{Engstrom et al.}
      {Engstrom, Yoder \& Crespi}{2016}]{EYC16}
      Engstrom T.A., Yoder N.C., Crespi V.H., 2016, ApJ, 818, 183

\bibitem[\protect\citeauthoryear{Chamel \& Fantina}{2016}]{CF16}
	    Chamel N., Fantina A.F., 2016, Phys.\ Rev.\ C, 94, 065802
	    
\bibitem[\protect\citeauthoryear{Fuchs}{1936}]{F36}
      Fuchs K., 1936, Proc.\ Roy.\ Soc.\ Lond., 153, 622

\bibitem[\protect\citeauthoryear{Haensel, Potekhin \& Yakovlev}{2007}]
      {HPY07}
      Haensel P., Potekhin A.Y., Yakovlev D.G., 2007, 
      Neutron Stars 1: Equation of State and Structure. Springer, 
      New York 

\bibitem[\protect\citeauthoryear{Horowitz \& Kadau}{2009}]{HK09}
      Horowitz C.J., Kadau K., 2009, Phys.\ Rev.\ Lett., 102, 191102

\bibitem[\protect\citeauthoryear{Johnson-McDaniel \& Owen}{2013}]{JO13}
	    Johnson-McDaniel N.K., Owen B.J., 2013, Phys.\ Rev.\ D, 
      88, 044004 

\bibitem[\protect\citeauthoryear{Kaspi}{2010}]{K10}
      Kaspi V.M., 2010, Proc.\ Nat.\ Acad.\ Sci., 107, 7147 

\bibitem[\protect\citeauthoryear{Kozhberov \& Baiko}{2015}]{KB15}
      Kozhberov A.A., Baiko D.A., 2015, Ap\&SS, 359, 50

\bibitem[\protect\citeauthoryear{Landau \& Lifshitz}{1980}]{LL80}
      Landau L.D., Lifshitz E.M., 1980, Statistical Physics. Part I.
      Pergamon Press, Oxford

\bibitem[\protect\citeauthoryear{Nagai \& Fukuyama}{1982}]{NF82}
      Nagai T., Fukuyama H., 1982, J.\ Phys.\ Soc.\ Jpn., 51, 3431

\bibitem[\protect\citeauthoryear{Nagai \& Fukuyama}{1983}]{NF83}
      Nagai T., Fukuyama H., 1983, J.\ Phys.\ Soc.\ Jpn., 52, 44

\bibitem[\protect\citeauthoryear{Nagara et al.}
      {Nagara, Nagata \& Nakamura}{1987}]{NNN87}
      Nagara H., Nagata Y., Nakamura T., 1987, 
      Phys.\ Rev.\ A, 36, 1859

\bibitem[\protect\citeauthoryear{Olausen \& Kaspi}{2014}]{OK14}
      Olausen S.A., Kaspi V.M., 2014, ApJS, 212, 6 \\ 
      (http://www.physics.mcgill.ca/$\sim$pulsar/magnetar/main.html) 

\bibitem[\protect\citeauthoryear{Piro}{2005}]{P05}
      Piro A., 2005, ApJ, 634, L153

\bibitem[\protect\citeauthoryear{Potekhin \& Yakovlev}{2012}]{PY12}
      Potekhin A.Y., Yakovlev D.G., 2012, Phys.\ Rev.\ C, 85, 039801

\bibitem[\protect\citeauthoryear{Schmidt et al.}{1997}]{SZRT97}
      Schmidt P., Zwicknagel G., Reinhard P.-G., Toepffer C., 1997,  
      Phys.\ Rev.\ E, 56, 7310

\bibitem[\protect\citeauthoryear{Sotani}{2016}]{S16}
      Sotani H., 2016, Phys.\ Rev.\ D, 93, 044059

\bibitem[\protect\citeauthoryear{Ushomirsky et al.}
      {Ushomirsky, Cutler \& Bildsten}{2000}]{UCB00}
      Ushomirsky G., Cutler C., Bildsten L., 2000, MNRAS, 319, 902

      
\end{thebibliography}
\end{document}